\documentclass[aps,prl,twocolumn,showpacs,superscriptaddress]{revtex4}   
\usepackage{graphicx}  
\usepackage{dcolumn}   
\usepackage{bm}        
\usepackage{amssymb}   
\usepackage{verbatim}  
\usepackage{epstopdf}
\usepackage{color}
\usepackage{soul}
\usepackage{amsmath}
\usepackage{subfigure}
\usepackage{color,colordvi,graphicx,pstcol}

\newcommand{\ket}[1]{\ensuremath{\left| #1 \right\rangle}}
\newcommand{\bra}[1]{\ensuremath{\left\langle #1 \right|}}

\newcommand{\beq}{\begin{equation}}
\newcommand{\eeq}{\end{equation}}
\newcommand{\bea}{\begin{eqnarray}}
\newcommand{\eea}{\end{eqnarray}}

\newcommand{\commentout}[1]{{}}

\usepackage[T1]{fontenc}
\usepackage{graphicx}
\usepackage{multirow,hhline,dcolumn} 
\usepackage[utf8]{inputenc}
\usepackage[normalem]{ulem}
\usepackage{setspace, mathtools}
\usepackage{amsmath, amsfonts, amssymb, physics, braket, mathrsfs}
\usepackage{indentfirst}
\definecolor{codegreen}{rgb}{0,0.6,0}
\definecolor{codegray}{rgb}{0.5,0.5,0.5}
\definecolor{codepurple}{rgb}{0.58,0,0.82}
\definecolor{backcolor}{rgb}{0.95,0.95,0.92}

\begin{document}

\title{Using Variational Eigensolvers on Low-End Hardware to Find the Ground State Energy of Simple Molecules}
\author{T. Powers}
\affiliation{Department of Physics, University of Massachusetts,
Dartmouth, MA 02747}
\author{R. M. Rajapakse}
\affiliation{Department of Physics, University of Massachusetts,
Dartmouth, MA 02747}
\affiliation{Department of Physics, University of Connecticut,
Storrs, CT 06269}

\date{\today}

\begin{abstract}
Key properties of physical systems can be described by the eigenvalues of matrices that represent the system. Computational algorithms that determine the eigenvalues of these matrices exist, but they generally suffer from a loss of performance as the matrix grows in size. This process can be expanded to quantum computation to find the eigenvalues with better performance than the classical algorithms. One application of such an eigenvalue solver is to determine energy levels of a molecule given a matrix representation of its Hamiltonian using the variational principle. Using a variational quantum eigensolver, we determine the ground state energies of different molecules. We focus on the choice of optimization strategy for a Qiskit simulator on low-end hardware. The benefits of several different optimizers were weighed in terms of accuracy in comparison to an analytic classical solution as well as code efficiency.
\end{abstract}


\maketitle

\section{Introduction}

     The Variational Quantum Eigensolver (VQE) is a method which uses the variational principle from quantum mechanics to estimate the minimum eigenvalue of a Hermitian matrix $H$ \cite{Qiskit-Textbook}. VQE is fairly simple to execute and is naturally resistant to error accumulation. Consequently, it is a candidate to be one of the first algorithms to practically outperform a classical computer \cite{McClean_Romero_Babbush_Aspuru-Guzik_2016} and yet researchers are still trying to find ways to enhance the process \cite{Jattana_Jin_De_Raedt_Michielsen_2023}. The algorithm estimates an eigenvalue $\lambda_\theta$ with corresponding eigenstate $\ket{\psi(\theta)}$, which is an upper bound for the minimum eigenvalue $\lambda_{min}$. This is described by the relation
    
    \begin{equation*}
        \lambda_{min} \leq \lambda_{\theta} = \bra{\psi(\theta)} H \ket{\psi(\theta)}.
    \end{equation*}
    
    Once this eigenstate is obtained, it is iteratively updated to minimize the expectation value of the matrix $\bra{\psi(\theta)} H \ket{\psi(\theta)}$. When this is minimized, it will be an approximation for the eigenvalue $\lambda_{min}$. 
    
    The procedure has several steps. First, an arbitrary initial state $\ket{\psi}$, an ansatz, is operated on by a parameterized circuit or variational form $U(\theta)$ or to obtain an initial estimate for $\ket{\psi(\theta)}$. The ansatz should approximate the eigenstate $\ket{\psi_{min}}$, whose associated eigenvalue is the desired $\lambda_{min}$. In some contexts, it can be shown \cite{Arrazola_2022} that initial state preparation can be performed in terms of Givens rotations, or rotations in a 2-dimensional plane.
    
    For the relevant problem and hardware being used, an appropriate optimizer is chosen and applied to $\ket{\psi(\theta)}$ to optimize the eigenstate. To preserve system resources and improve efficiency, the number of free parameters in the variational form should be minimized. In the case of a 1-qubit variational form, the operator
    
    \begin{equation*}
        U3(\theta, \phi, \lambda) =
        \begin{bmatrix*}
            \cos \frac{\theta}{2} & -e^{i \lambda} \sin \frac{\theta}{2} \\
            e^{i \phi} \sin \frac{\theta}{2} & e^{i \lambda + i \phi} \cos \frac{\theta}{2}
        \end{bmatrix*}
    \end{equation*}
    
    \noindent can be used to apply any single qubit transformation when the proper parameters are chosen, aside from a global phase. This form has only the three free parameters $\theta, \phi,$ and $\lambda$, and can be optimized efficiently. However, systems often require more than a single qubit to be represented on a quantum device.
    
    Several kinds of optimization strategies can be employed. One of these is called ``stochastic gradient descent,” where parameters are updated one at a time in a manner which causes the objective function to fluctuate \cite{Ruder_2016}. However, this method generally evaluates the circuit an expensive number of times and can often result in getting stuck at only local optima. For these reasons, it is not recommended for VQE on noisy devices \cite{Qiskit-Textbook}. Some other optimizers make use of random perturbation, where the parameters are updated in a random fashion to reduce the impact of noise.
    
    After the optimizer is applied for enough iterations, the estimated energy expectation value $\lambda_{\theta}$ should be a good approximation of the minimum eigenvalue $\lambda_{min}$.  When the algorithm is given a Hamiltonian matrix, this eigenvalue is a representation of the system’s ground state energy.

\section{Initial Code \& Modifications}
    
    Some code was taken from the ``Simulating Molecules using VQE'' section 
    of the Qiskit online textbook \cite{Qiskit-Textbook} to serve as a starting point. Using VQE with the Sequential Least Squares Programming (SLSQP) optimizer, the code calculates the ground state energy of Lithium Hydride at several atomic distances using a quantum simulator powered by the Qiskit developer kit. The same energy is computed with a classical solver for comparison.
    
    The code first initializes the statevector simulator, computes the list of atomic distances which will be used for the simulations, and creates empty lists to store both the VQE-computed energies and the classically-calculated energies.
    
    Next it runs an initialization routine. This takes a defined molecule function, which itself is a function of atomic distance, containing the elements and positions of the atoms making up the molecule.  The lithium atom is positioned at the origin coordinate $(0,0,0)$ and the Hydrogen atom is placed on the X-axis some distance away, which is the argument of the function. For molecules of more than two atoms, variational methods have been conceived \cite{Delgado_2021} which determine the optimal angle geometry for calculating minimum energies. For molecules of only two atoms, though, there are no angles to consider.
    
    The code then uses this molecule function for the first atomic distance in the series to initialize a driver, which processes the properties of the defined molecule in the STO-3G basis. This minimal basis set uses 3 Gaussian-Type Orbitals (GTO) to approximate the accurate yet expensive Slater-Type orbitals (STO) in a manner that is less resource-exhaustive but not greatly accurate \cite{STO}. The number of particles and spin orbitals is collected from this driver. The orbitals are removed, the Hamiltonian is set as a sum of Pauli operators, and two-qubit reduction is performed. This reduction is done along with freezing the core to reduce the total number of qubits required \cite{TwoQubitReduction}.
    
    The result is calculated classically with an eigensolver function, then by VQE using the SLSQP optimizer with a Hartree-Fock ansatz and the UCCSD variational form. The Hartree-Fock method ``approximates the wave function [of multiple electrons] in terms of single-electron wave functions'' \cite{hartreeFock}, and the UCCSD variational form is a unitary form of the classical coupled-cluster (CC) variational form for single- and double-excitation wavefunctions \cite{Barkoutsos_2018}. The results are then calculated for the remaining atomic distances before being plotted.
    
    Once the initial code was in place, a few major changes were made for the main body of this work. First, two new optimizers were introduced to the code. This was done so that both accuracy and effiency could be compared in an effort to choose the best optimizer for the available hardware. These new optimizers were the Constrained Optimization by Linear Approximation (COBYLA) and Simultaneous Perturbation Stochastic Approximation (SPSA) optimizers, both supported natively by the Qiskit libraries.
    
    Besides adding new optimizers, different molecules were examined altogether. These new molecules were Hydrogen gas (H$_2$), Hydrogen Fluoride (HF), Hydrogen Chloride (HCl), and Imidogen (NH). Each of these molecules had their energies calculated by VQE using each of the three optimizers. A timer was also added to the code to record the runtime for each of the trials.

\section{Accuracy}

\begin{figure}[hbtp!]
        \begin{center}
            \includegraphics[scale = 0.6]{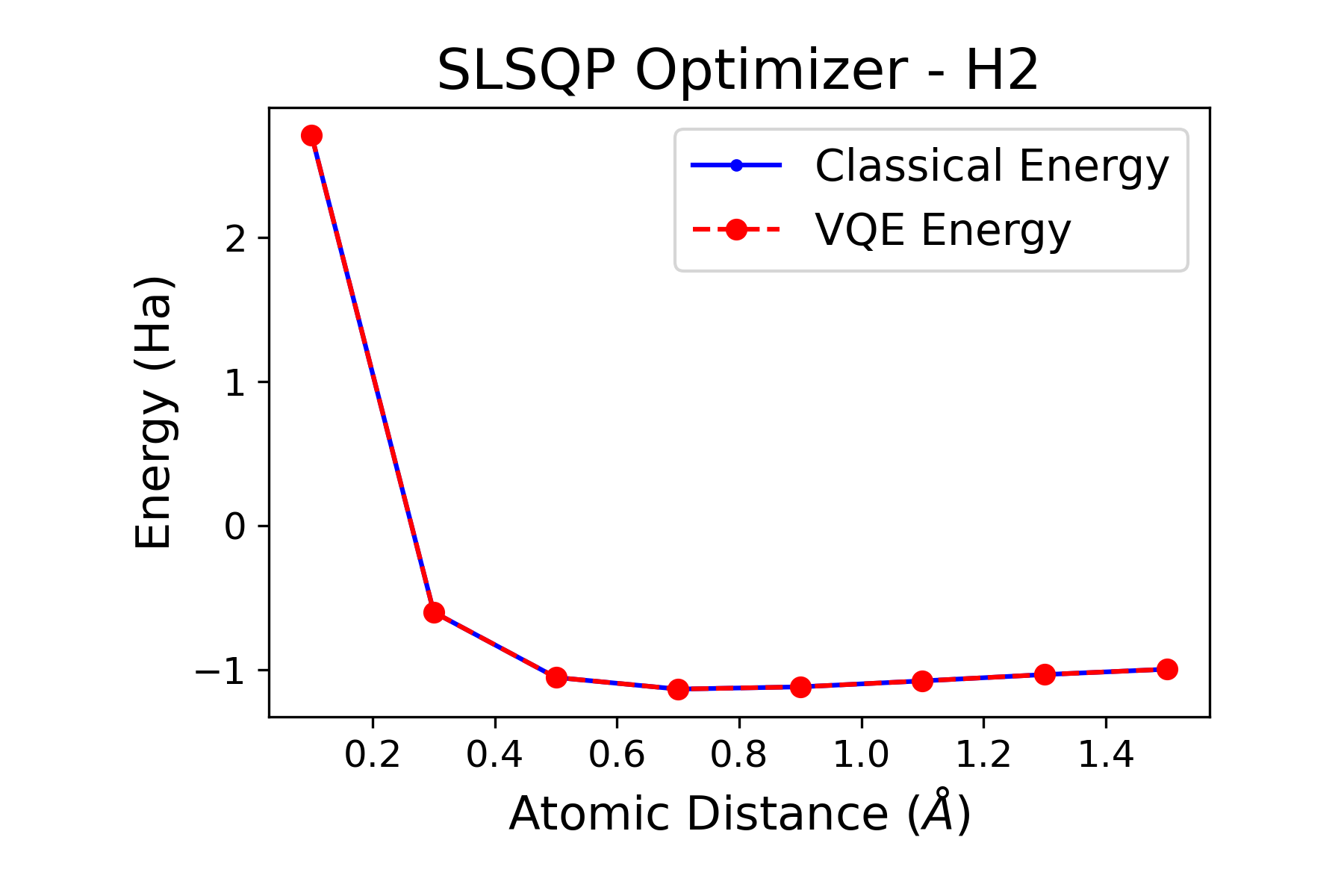}
            \caption[Energy vs Atomic Distance plot for Hydrogen Gas (H$_2$) with the SLSQP optimizer.]{Energy vs Atomic Distance plot for Hydrogen Gas (H$_2$) with the SLSQP optimizer. The minimum energy calculated was $-1.136$ Ha at a distance of $0.7$ Angstroms.}
            \label{SLSQP H2}
        \end{center}
    \end{figure}

\begin{figure}[hbtp!]
        \begin{center}
            \includegraphics[scale = 0.6]{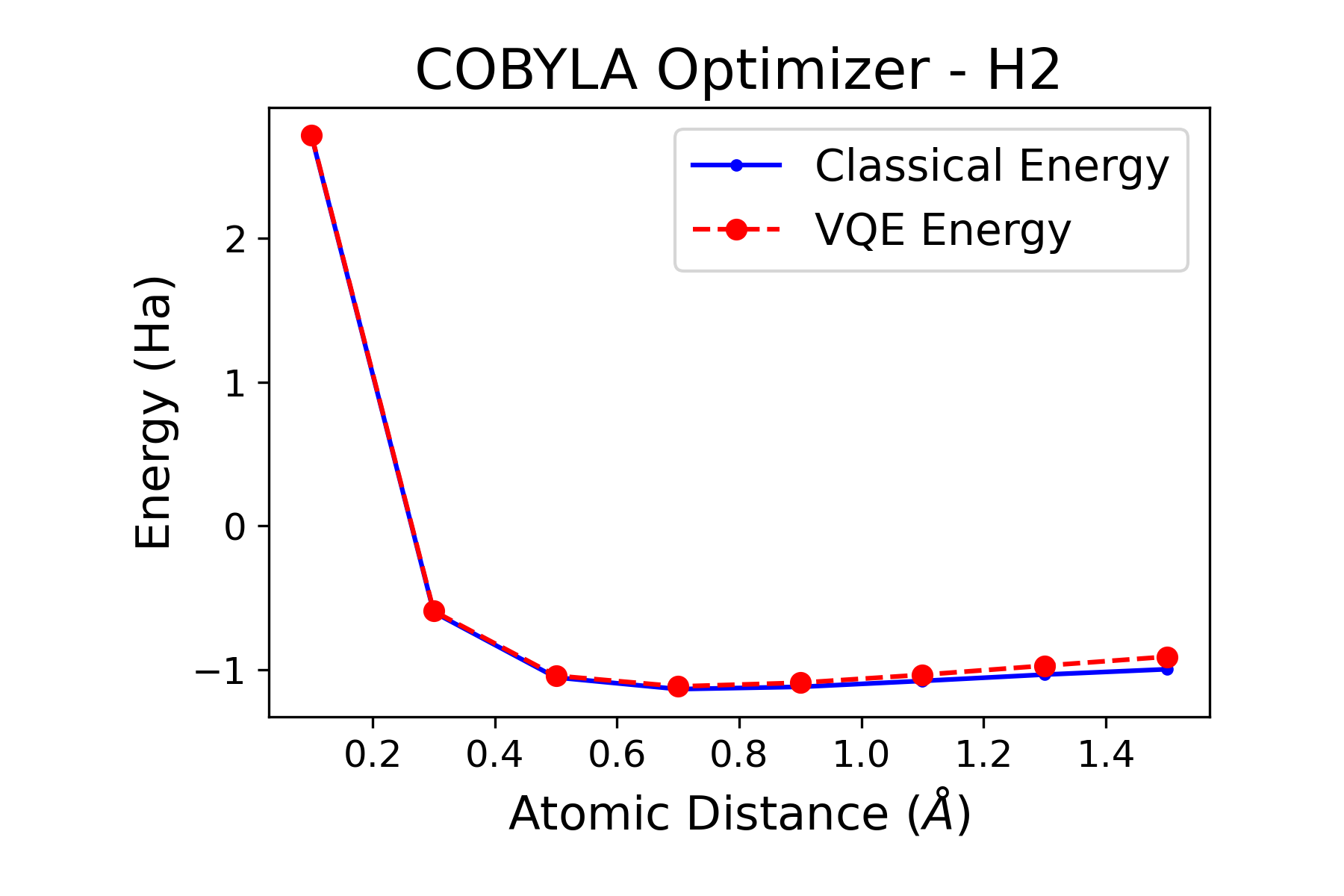}
            \caption[Energy vs Atomic Distance plot for Hydrogen Gas (H$_2$) with the COBYLA optimizer.]{Energy vs Atomic Distance plot for Hydrogen Gas (H$_2$) with the COBYLA optimizer. The minimum energy calculated was $-1.117$ Ha at a distance of $0.7$ Angstroms.}
            \label{COBYLA H2}
        \end{center}
    \end{figure}

\begin{figure}[hbtp!]
        \begin{center}
            \includegraphics[scale = 0.6]{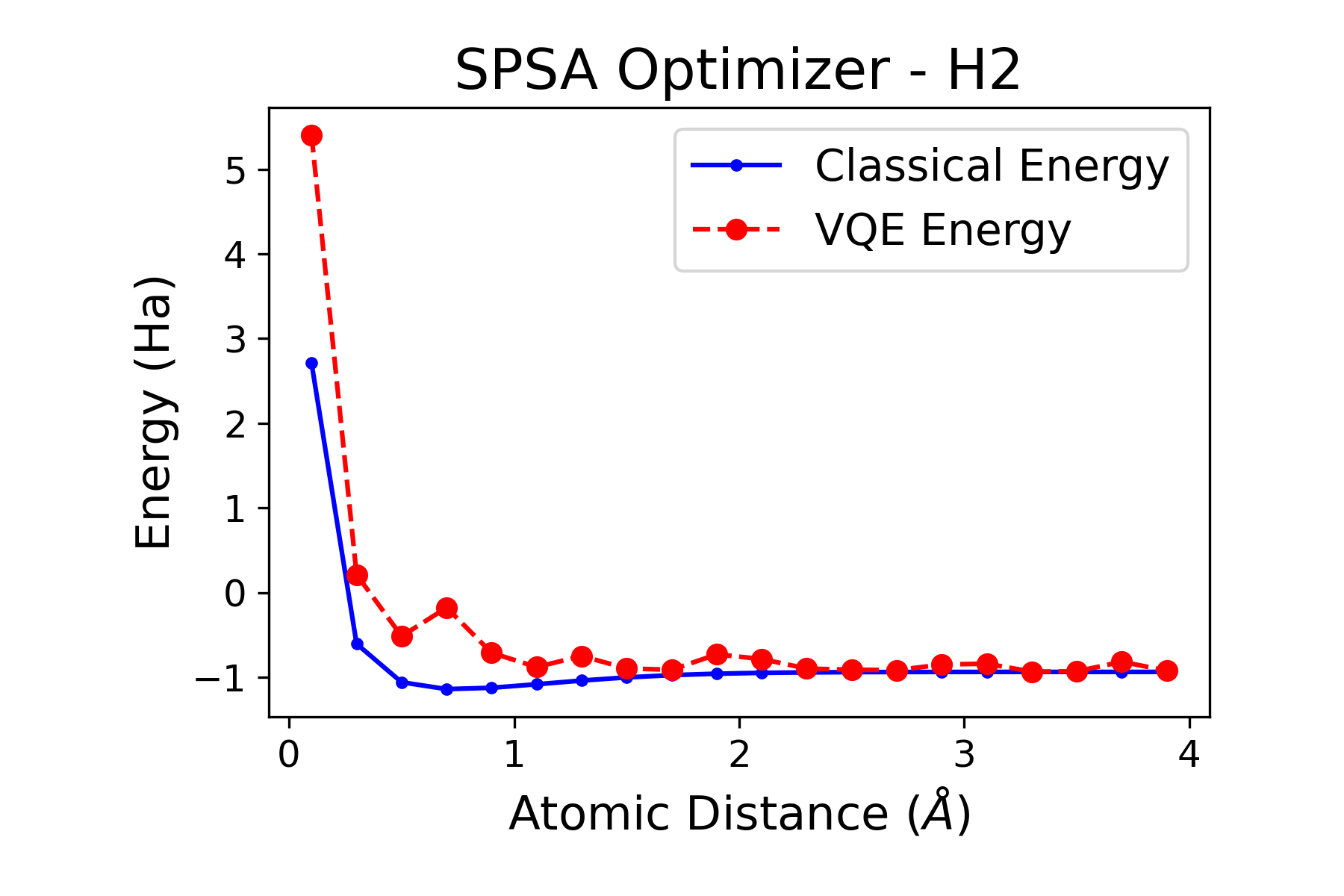}
            \caption[Energy vs Atomic Distance plot for Hydrogen Gas (H$_2$) with the SPSA optimizer.]{Energy vs Atomic Distance plot for Hydrogen Gas (H$_2$) with the SPSA optimizer. For this particular run, the minimum energy calculated was $-0.930$ Ha at a distance of $3.3$ Angstroms.}
            \label{SPSA H2}
        \end{center}
    \end{figure}

\begin{figure}[hbtp!]
        \begin{center}
            \includegraphics[scale = 0.6]{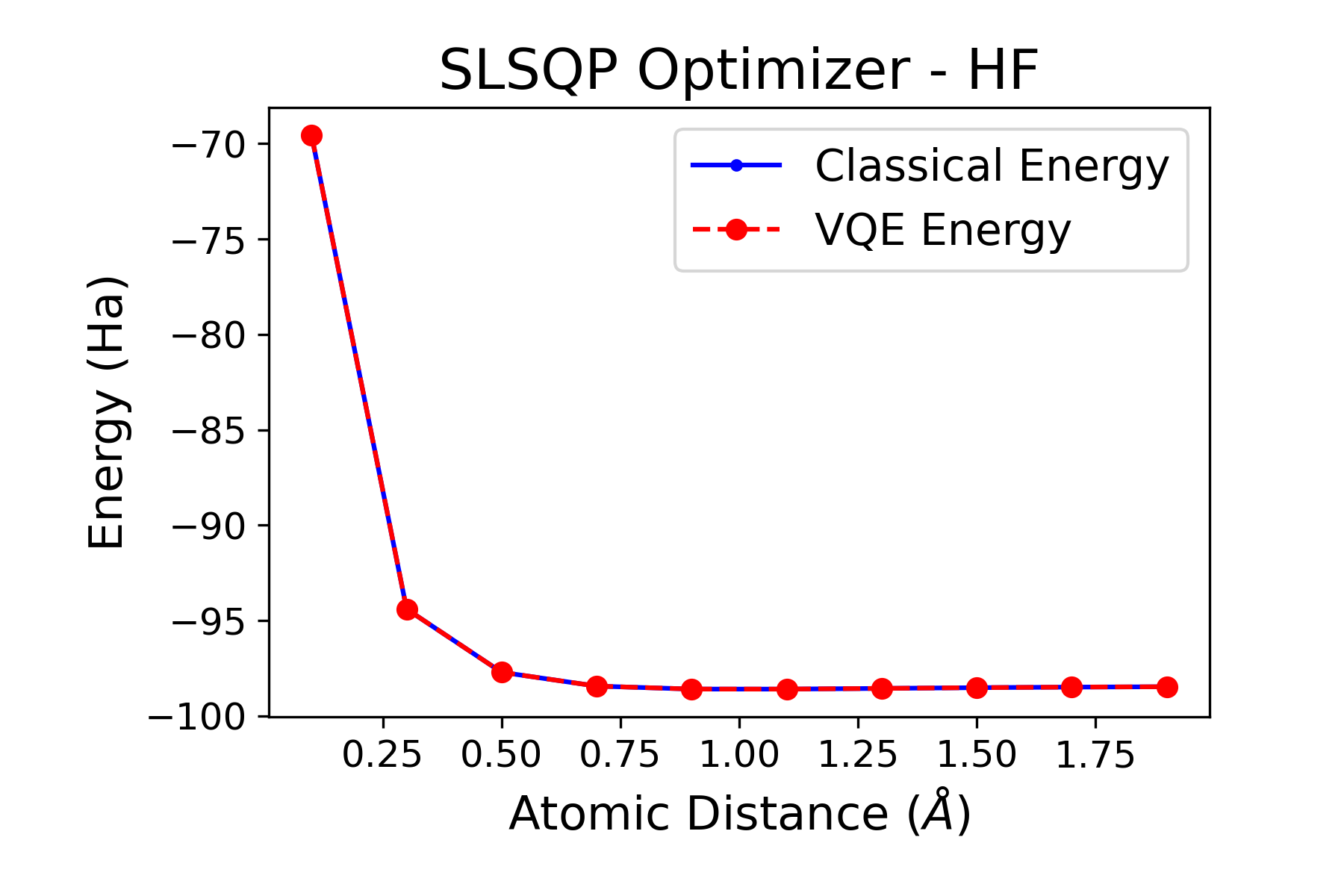}
            \caption[Energy vs Atomic Distance plot for Hydrogen Fluoride (HF) with the SLSQP optimizer.]{Energy vs Atomic Distance plot for Hydrogen Fluoride (HF) with the SLSQP optimizer. The minimum energy calculated was $-98.595$ Ha at a distance of $1.1$ Angstroms.}
            \label{SLSQP HF}
        \end{center}
    \end{figure}

\begin{figure}[hbtp!]
        \begin{center}
            \includegraphics[scale = 0.6]{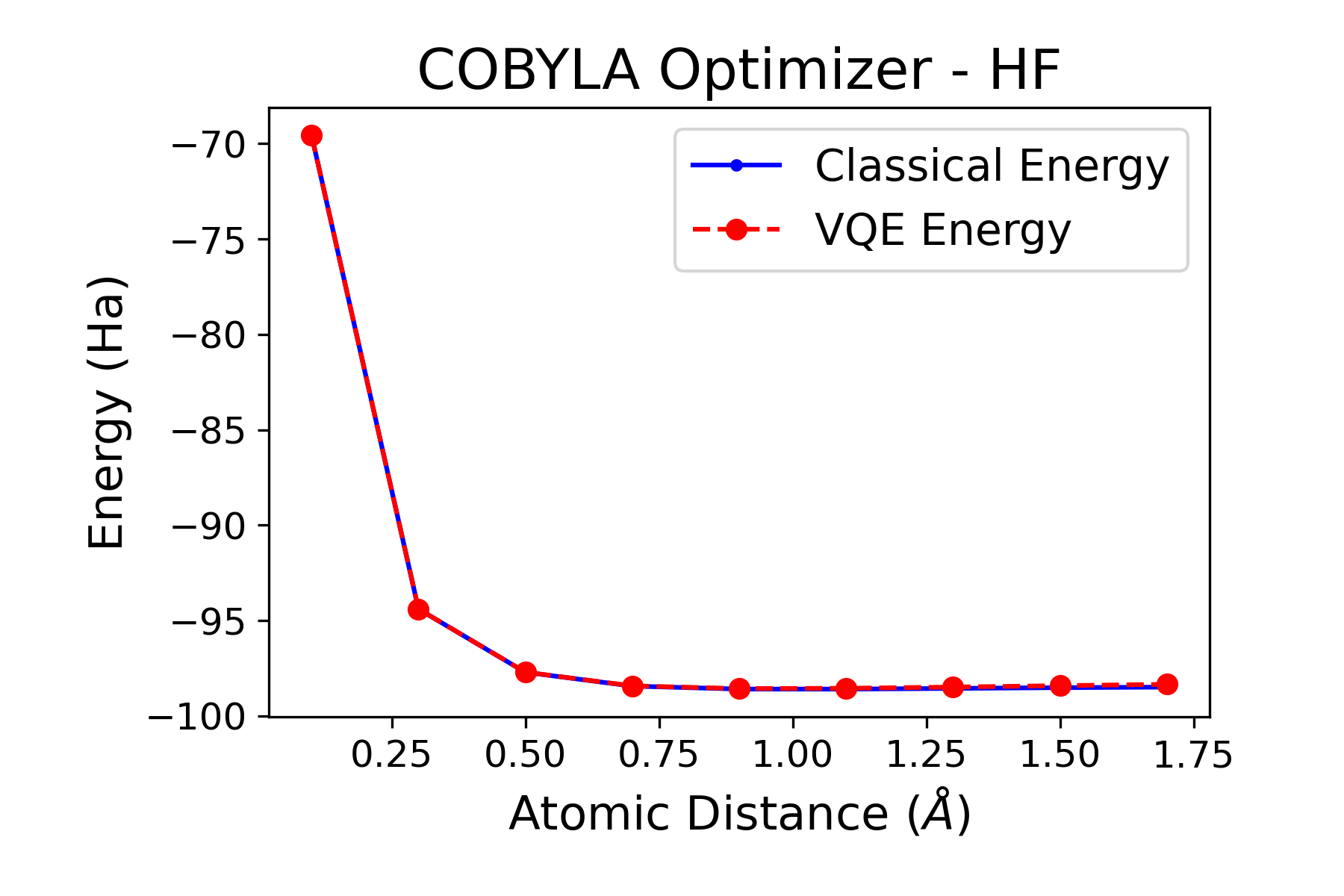}
            \caption[Energy vs Atomic Distance plot for Hydrogen Fluoride (HF) with the COBYLA optimizer.]{Energy vs Atomic Distance plot for Hydrogen Fluoride (HF) with the COBYLA optimizer. The minimum energy calculated was $-98.568$ Ha at a distance of $0.9$ Angstroms.}
            \label{COBYLA HF}
        \end{center}
    \end{figure}

\begin{figure}[hbtp!]
        \begin{center}
            \includegraphics[scale = 0.6]{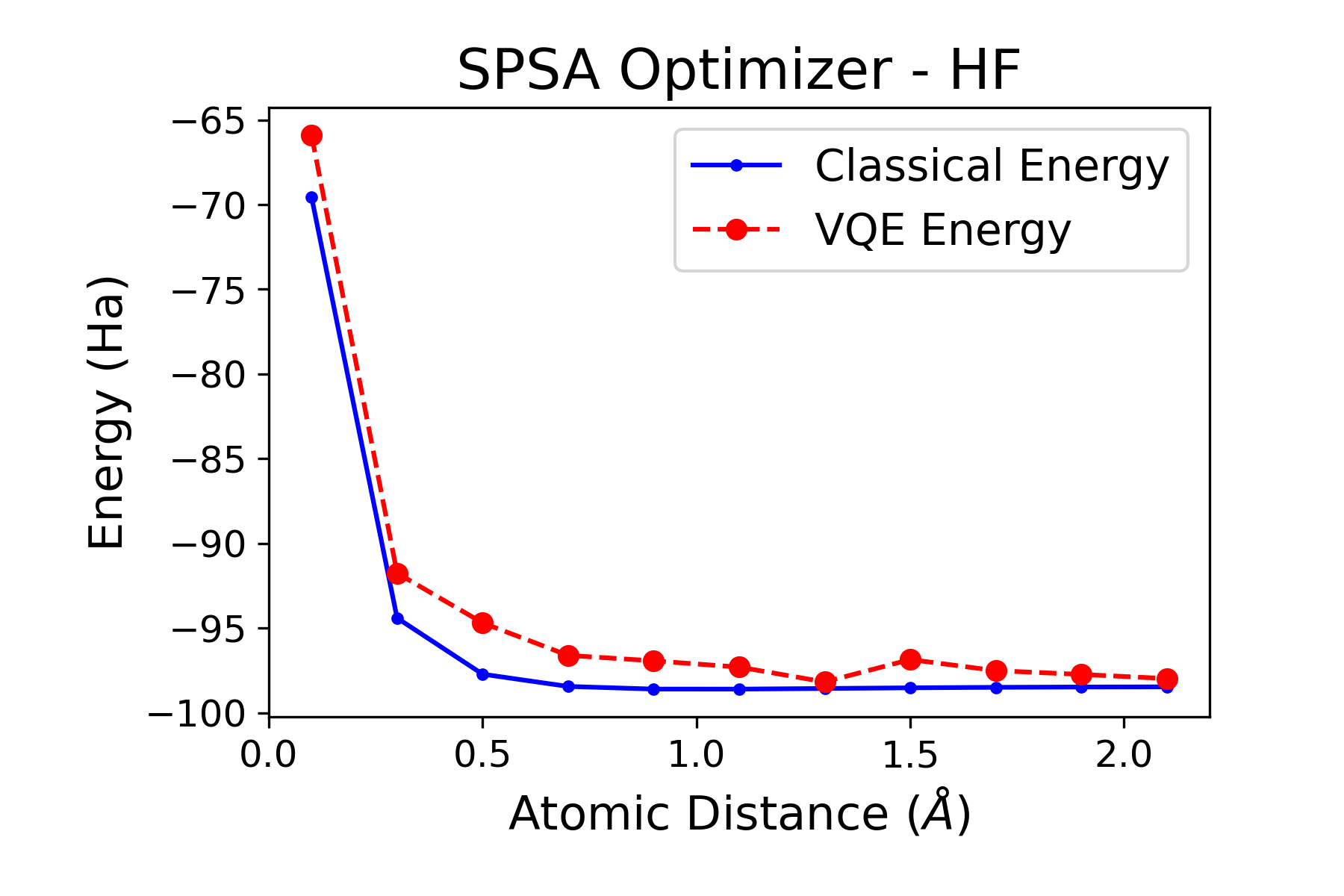}
            \caption[Energy vs Atomic Distance plot for Hydrogen Fluoride (HF) with the SPSA optimizer.]{Energy vs Atomic Distance plot for Hydrogen Fluoride (HF) with the SPSA optimizer. For this particular run, the minimum energy calculated was $-98.172$ Ha at a distance of $1.3$ Angstroms.}
            \label{SPSA HF}
        \end{center}
    \end{figure}

\begin{figure}[hbtp!]
        \begin{center}
            \includegraphics[scale = 0.6]{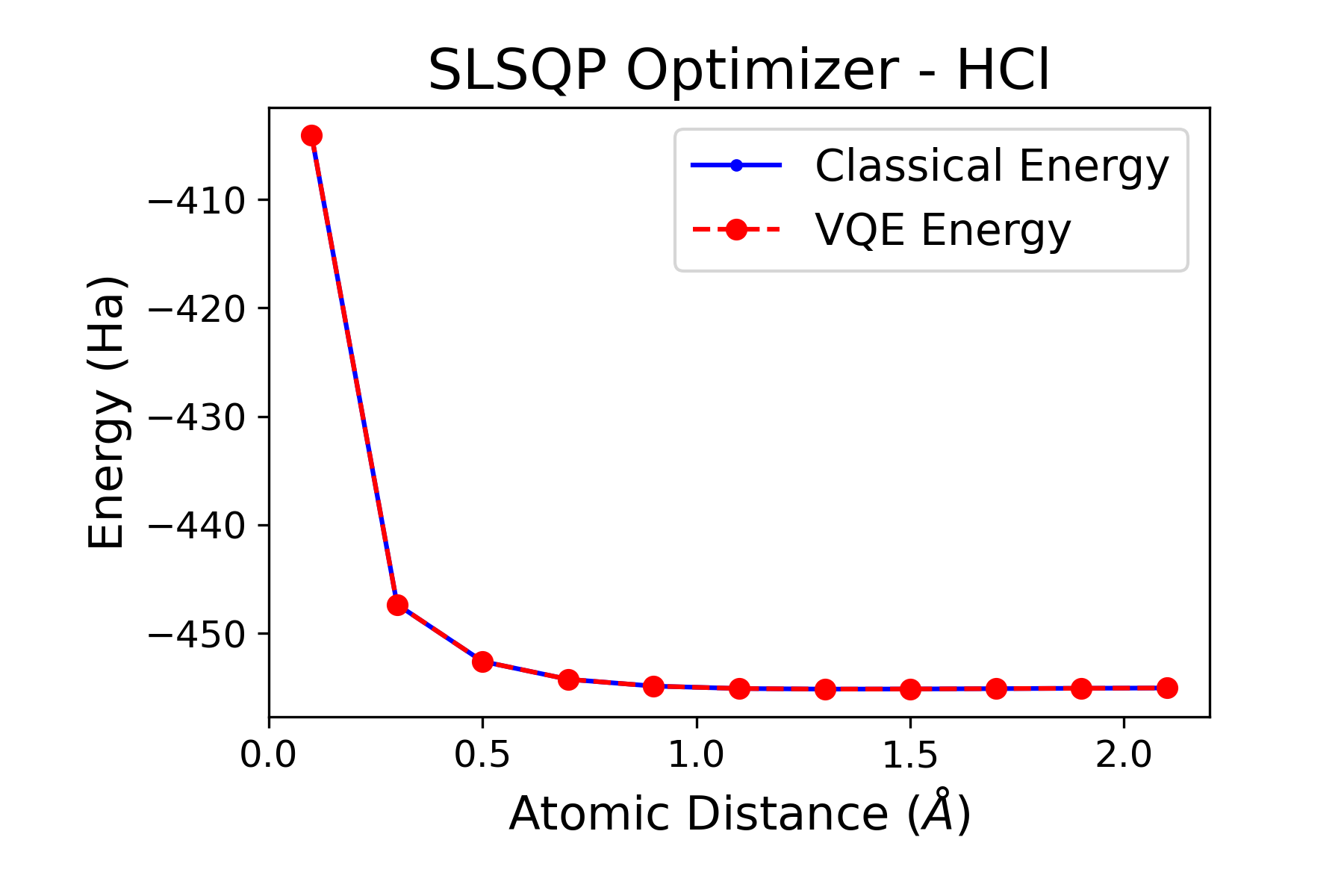}
            \caption[Energy vs Atomic Distance plot for Hydrogen Chloride (HCl) with the SLSQP optimizer.]{Energy vs Atomic Distance plot for Hydrogen Chloride (HCl) with the SLSQP optimizer. The minimum energy calculated was $-455.156$ Ha at a distance of $1.3$ Angstroms.}
            \label{SLSQP HCl}
        \end{center}
    \end{figure}

\begin{figure}[hbtp!]
        \begin{center}
            \includegraphics[scale = 0.6]{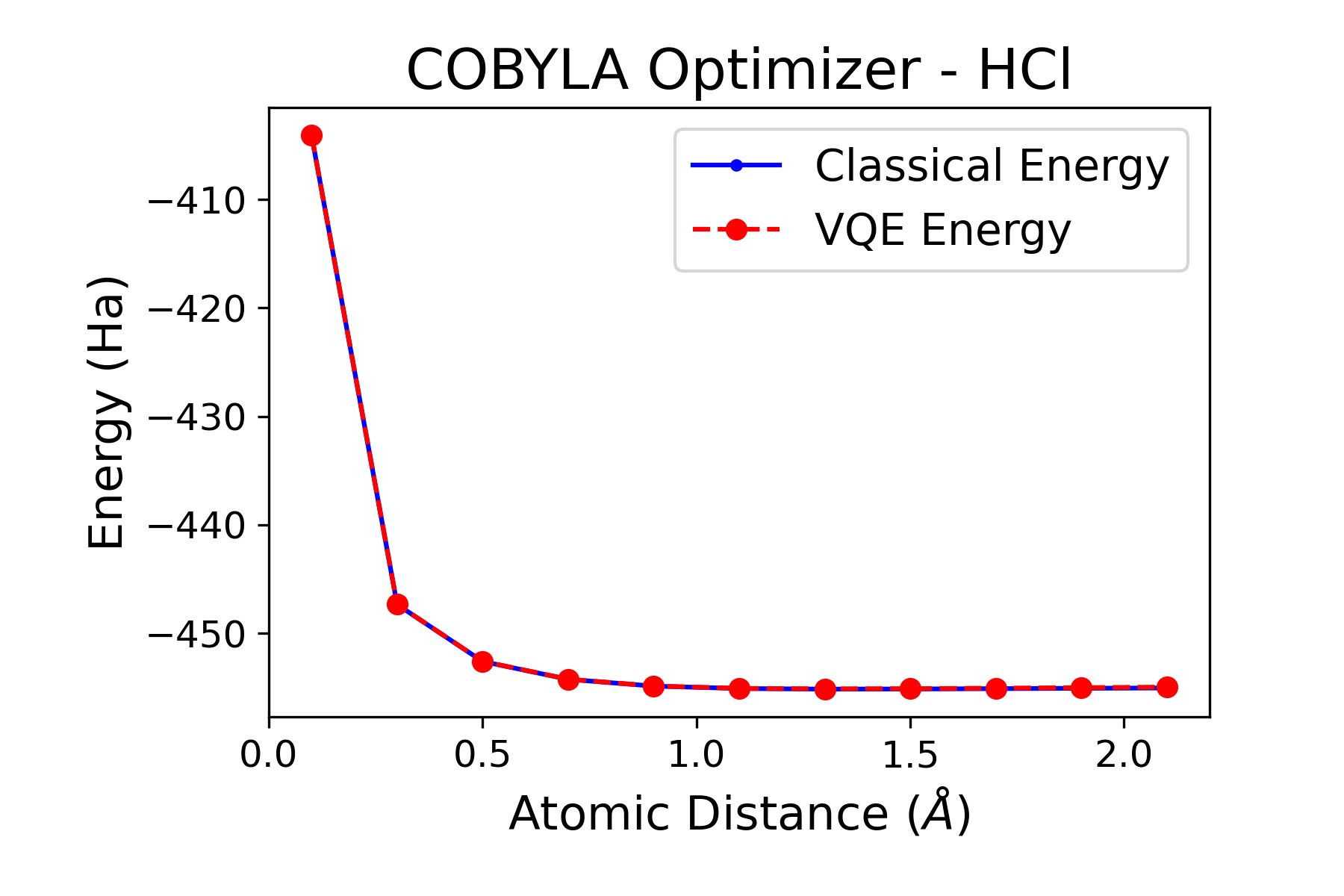}
            \caption[Energy vs Atomic Distance plot for Hydrogen Chloride (HCl) with the COBYLA optimizer.]{Energy vs Atomic Distance plot for Hydrogen Chloride (HCl) with the COBYLA optimizer. The minimum energy calculated was $-455.136$ Ha at a distance of $1.3$ Angstroms.}
            \label{COBYLA HCl}
        \end{center}
    \end{figure}

\begin{figure}[hbtp!]
        \begin{center}
            \includegraphics[scale = 0.6]{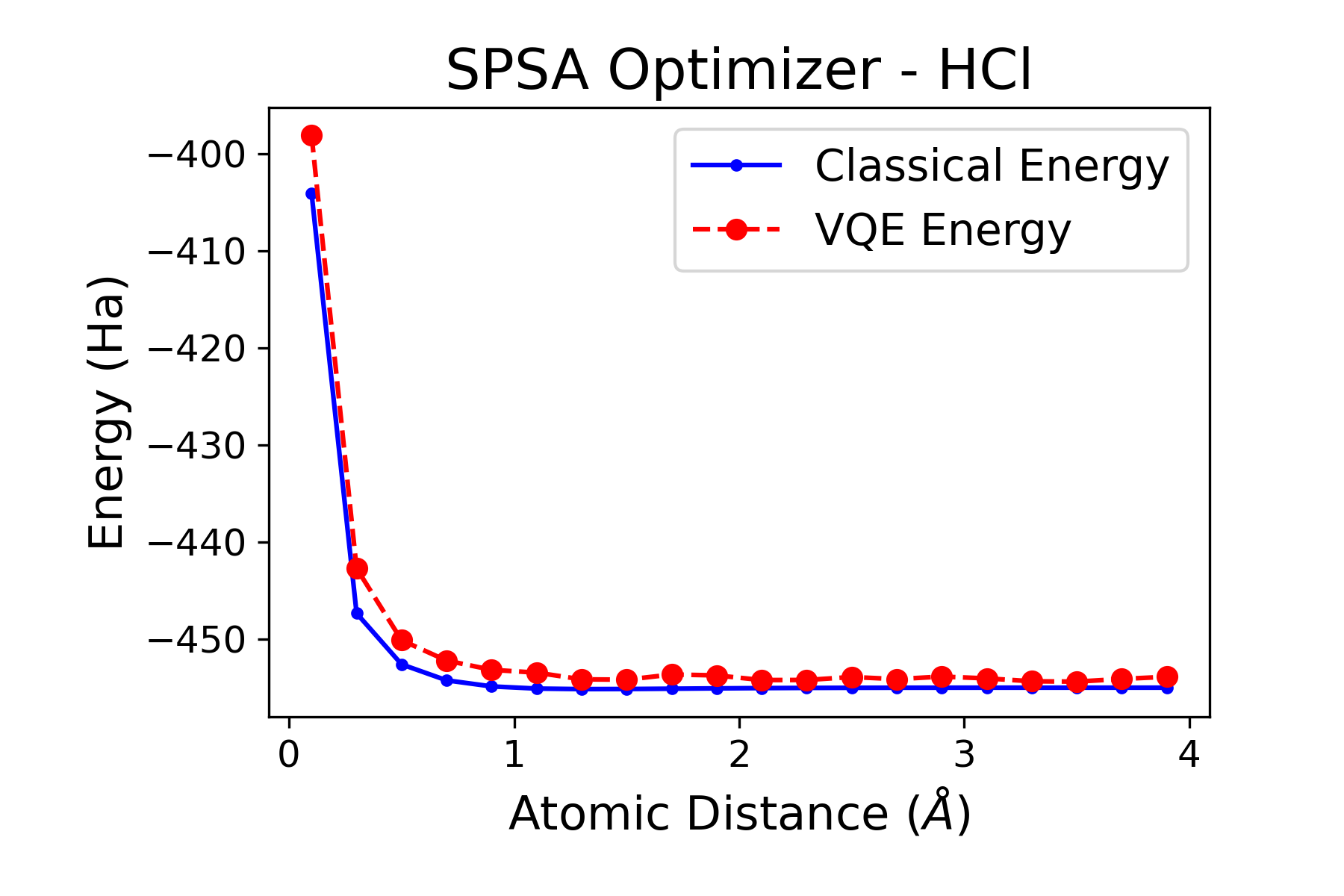}
            \caption[Energy vs Atomic Distance plot for Hydrogen Chloride (HCl) with the SPSA optimizer.]{Energy vs Atomic Distance plot for Hydrogen Chloride (HCl) with the SPSA optimizer. For this particular run, the minimum energy calculated was $-454.384$ Ha at a distance of $3.5$ Angstroms.}
            \label{SPSA HCl}
        \end{center}
    \end{figure}

\begin{figure}[hbtp!]
        \begin{center}
            \includegraphics[scale = 0.6]{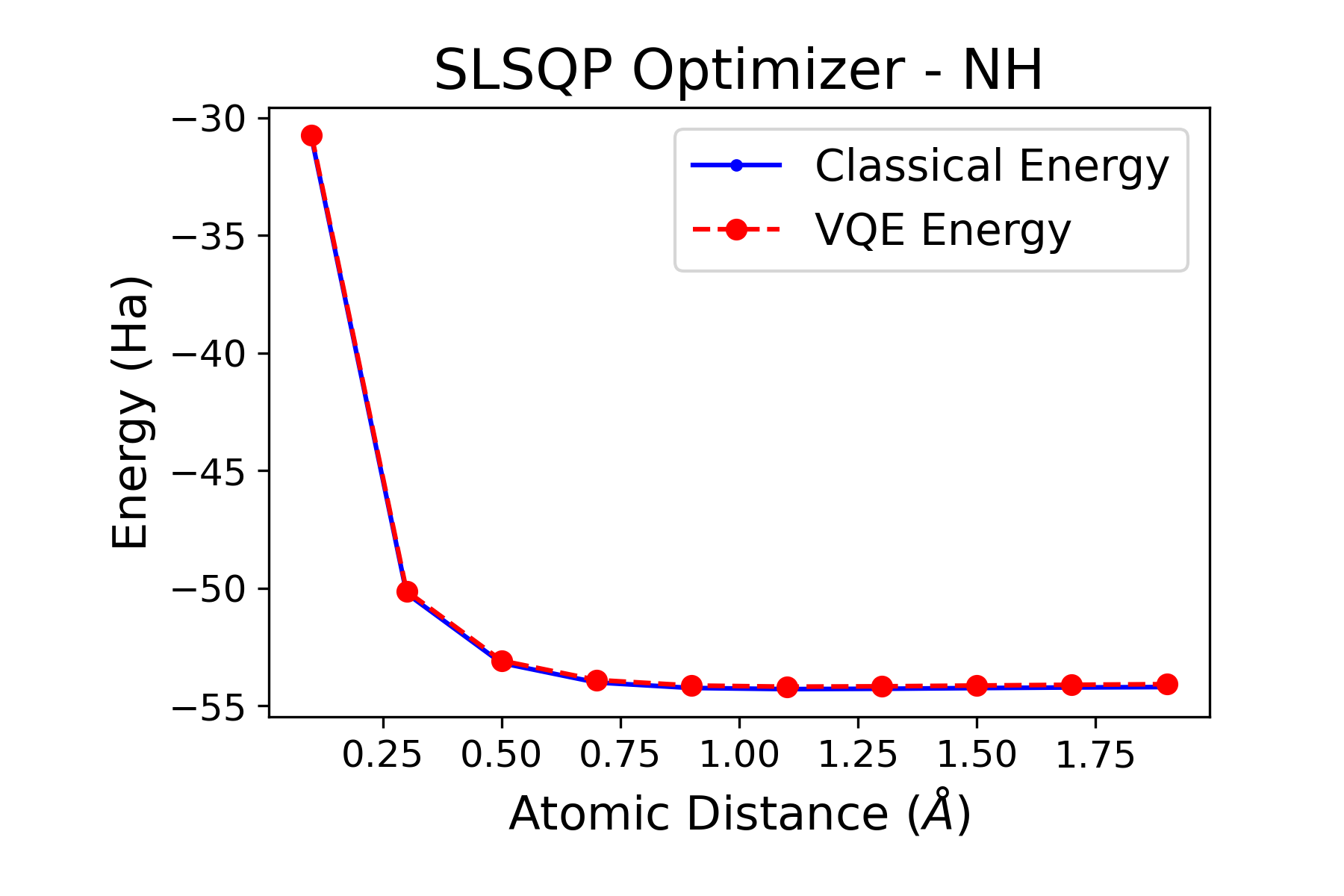}
            \caption[Energy vs Atomic Distance plot for Imidogen (NH) with the SLSQP optimizer.]{Energy vs Atomic Distance plot for Imidogen (NH) with the SLSQP optimizer. The minimum energy calculated was $-54.185$ Ha at a distance of $1.1$ Angstroms.}
            \label{SLSQP NH}
        \end{center}
    \end{figure}

\begin{figure}[hbtp!]
        \begin{center}
            \includegraphics[scale = 0.6]{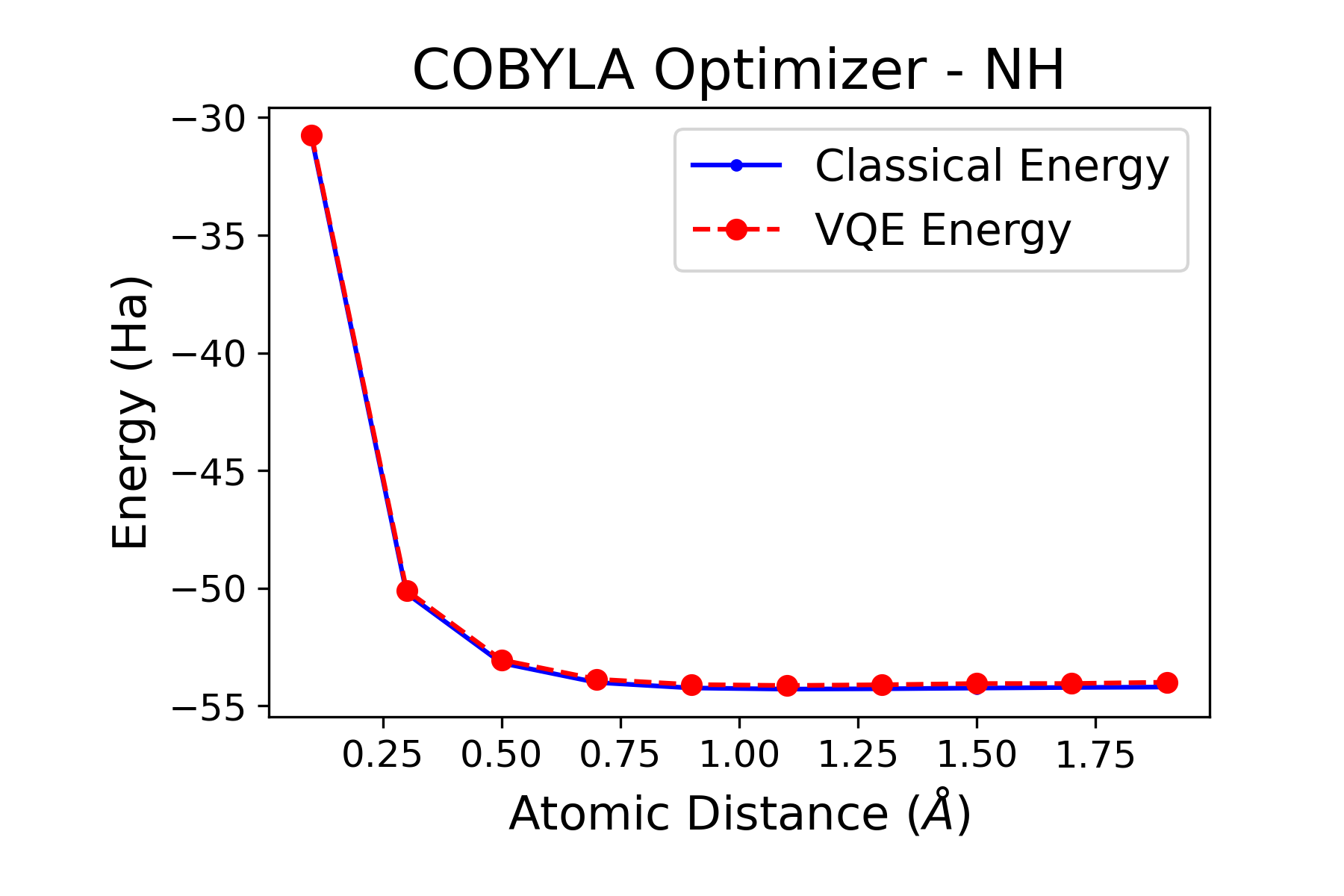}
            \caption[Energy vs Atomic Distance plot for Imidogen (NH) with the COBYLA optimizer.]{Energy vs Atomic Distance plot for Imidogen (NH) with the COBYLA optimizer. The minimum energy calculated was $-54.136$ Ha at a distance of $1.1$ Angstroms.}
            \label{COBYLA NH}
        \end{center}
    \end{figure}

\begin{figure}[hbtp!]
        \begin{center}
            \includegraphics[scale = 0.6]{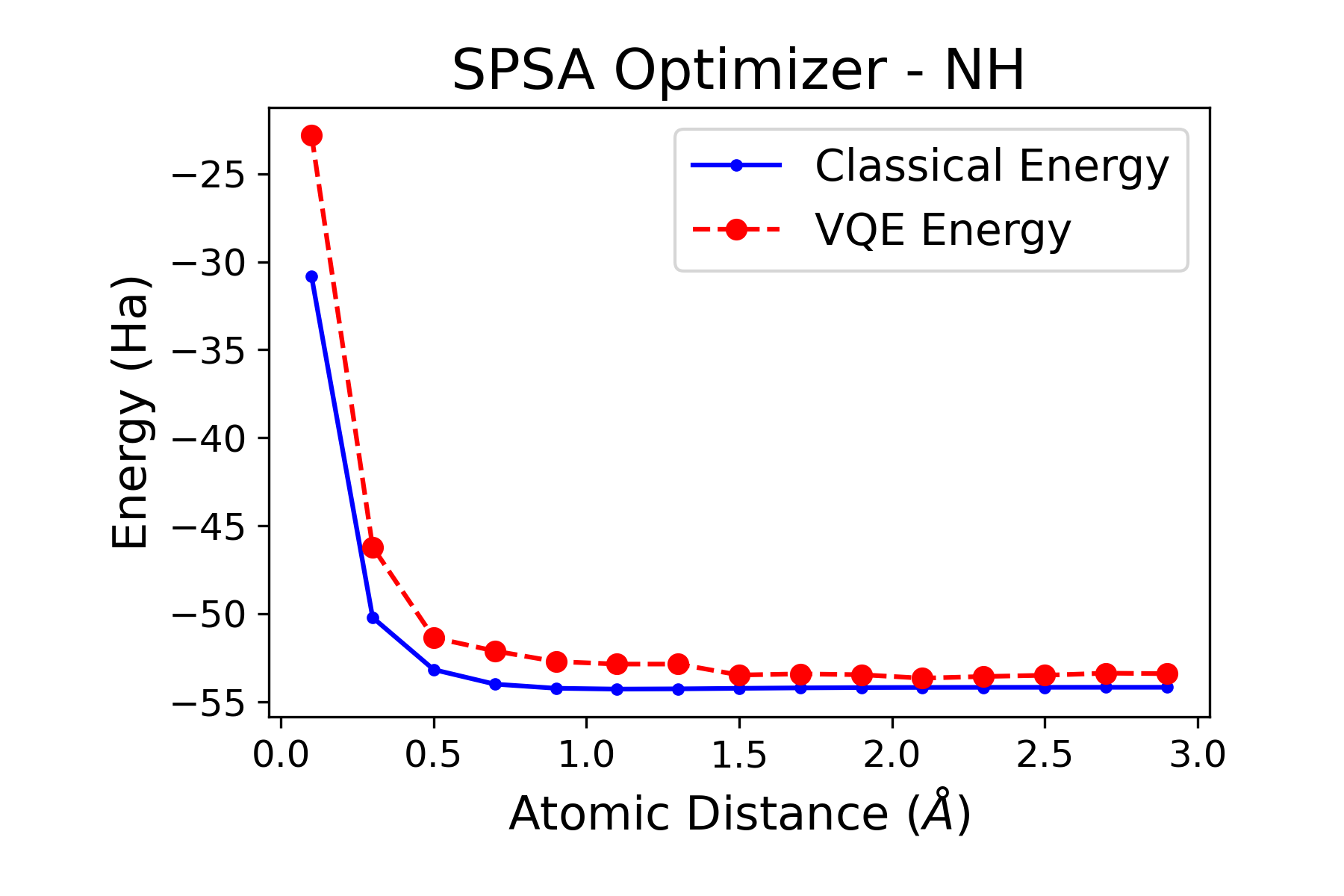}
            \caption[Energy vs Atomic Distance plot for Imidogen (NH) with the SPSA optimizer.]{Energy vs Atomic Distance plot for Imidogen (NH) with the SPSA optimizer. For this particular run, the minimum energy calculated was $-53.669$ Ha at a distance of $2.1$ Angstroms.}
            \label{SPSA NH}
        \end{center}
    \end{figure}

    Energy vs Atomic Distance plots were produced by the code using each of the different optimizers. The plots give a demonstration of the accuracy of the process. Comparing the VQE energy curves to the classical energy curves quickly illuminates the difference in the calculations at a given atomic distance. These plots are shown in Figures \ref{SLSQP H2} through \ref{SPSA NH}. It should be noted that the energies, measured in Hartrees, are calculated for atomic distances between $0.1$ and $4.0$ Angstroms, but most of the plots show a restricted domain to better present the minimum values of the energy curves. Also, the random nature of the SPSA optimizer causes the energy values to vary slightly between different runs of the code, and so the plots for this optimizer are not always easily reproducible without providing a seed manually.
    
    \newpage
    
    \newpage
    
    \newpage
    
    \newpage
    
    \newpage
    
    \newpage
    
    Generally, the SLSQP optimizer consistently produces the energy curve which most closely fits the classical curve of the three optimizers examined (Figs \ref{SLSQP H2}, \ref{SLSQP HF}, \ref{SLSQP HCl}, \ref{SLSQP NH}). For this reason, we consider the SLSQP optimizer to produce the most accurate results of these tested optimizers. Most of the time, the curve for the code run with the COBYLA optimizer is also quite accurate, though the curves often begin to diverge slightly after the minimum value has been reached (Figs \ref{COBYLA H2}, \ref{COBYLA HF}, \ref{COBYLA HCl}, \ref{COBYLA NH}). This is most apparent in the plot for Hydrogen Gas (H$_2$) using the COBYLA optimizer, but is also the case for the other molecules outside of the domain presented in the plots above. The VQE curves produced using the SPSA optimizer are consistently the least close to the classical curves (Figs \ref{SPSA H2}, \ref{SPSA HF}, \ref{SPSA HCl}, \ref{SPSA NH}). The difference between the curves is most noticeable for the case of Hydrogen Gas (H$_2$).

\section{Efficiency}

    In computation, minimizing runtime can be just as import as having accurate results. Often, waiting long periods of time for slight improvements in precision is not desirable. Tables \ref{Start Runtime Tables} through \ref{End Runtime Tables} contain the runtimes for the code for each of the molecules. Once again, these runtimes represent the calculations for atomic distances between $0.1$ and $4.0$ Angstroms.
    
    \begin{table}[htbp!]
        \caption{Runtimes for VQE code simulating H$_2$ molecule.}
        \label{Start Runtime Tables}
        \begin{center}
            \begin{tabular}{ || c | c | c | c || }
                \hhline{ |t: = = = = :t| } 
                Trial & SLSQP & COBYLA & SPSA \\
                 & (s) & (s) & (s) \\
                \hhline{ ||- - - -|| }      
                1 & 18.24 & 17.02 & 19.45  \\
                \hhline{ ||- - - -|| }
                2 & 19.73 & 17.60 & 19.53 \\
                \hhline{ ||- - - -|| }
                3 & 18.18 & 17.20 & 19.42 \\
                \hhline{ |b: = = = = :b| } 
            \end{tabular}
        \end{center}
    \end{table}
    
    \pagebreak
    
    \begin{table}[htbp!]
        \caption{Runtimes for VQE code simulating HF molecule.}
        \begin{center}
            \begin{tabular}{ || c | c | c | c || }
                \hhline{ |t: = = = = :t| } 
                Trial & SLSQP & COBYLA & SPSA \\
                & (m:s) & (m:s) & (m:s) \\
                \hhline{ ||- - - -|| }      
                1 & 10:09.13 & 02:50.63 & 06:04.99 \\
                \hhline{ ||- - - -|| }
                2 & 09:36.79 & 02:48.99 & 06:08.58 \\
                \hhline{ ||- - - -|| }
                3 & 09:41.35 & 02:52.70 & 06:11.98 \\
                \hhline{ |b: = = = = :b| } 
            \end{tabular}
        \end{center}
    \end{table}
    
    \begin{table}[htbp!]
        \caption{Runtimes for VQE code simulating HCl molecule.}
        \begin{center}
            \begin{tabular}{ || c | c | c | c || }
                \hhline{ |t: = = = = :t| } 
                Trial & SLSQP & COBYLA & SPSA \\
                & (m:s) & (m:s) & (m:s) \\
                \hhline{ ||- - - -|| }
                1 & 10:12.08 & 02:57.69 & 06:12.90 \\
                \hhline{ ||- - - -|| }
                2 & 10:11.25 & 02:54.94 & 06:19.45 \\
                \hhline{ ||- - - -|| }
                3 & 09:42.63 & 02:59.25 & 06:23.85 \\
                \hhline{ |b: = = = = :b| } 
            \end{tabular}
        \end{center}
    \end{table}
    
    \begin{table}[htbp!]
        \caption{Runtimes for VQE code simulating NH molecule.}
        \label{End Runtime Tables}
        \begin{center}
            \begin{tabular}{ || c | c | c | c || }
                \hhline{ |t: = = = = :t| } 
                Trial & SLSQP & COBYLA & SPSA \\
                & (m:s) & (m:s) & (m:s) \\
                \hhline{ ||- - - -|| }      
                1 & 43:56.93 & 05:12.78 & 13:33.42 \\
                \hhline{ ||- - - -|| }
                2 & 44:01.80 & 05:09.54 & 13:22.93 \\
                \hhline{ ||- - - -|| }
                3 & 43:48.15 & 05:12.29 & 13:22.46 \\
                \hhline{ |b: = = = = :b| } 
            \end{tabular}
        \end{center}
    \end{table}
    
\begin{figure}[hbtp!]
        \begin{center}
            \includegraphics[scale = 0.6]{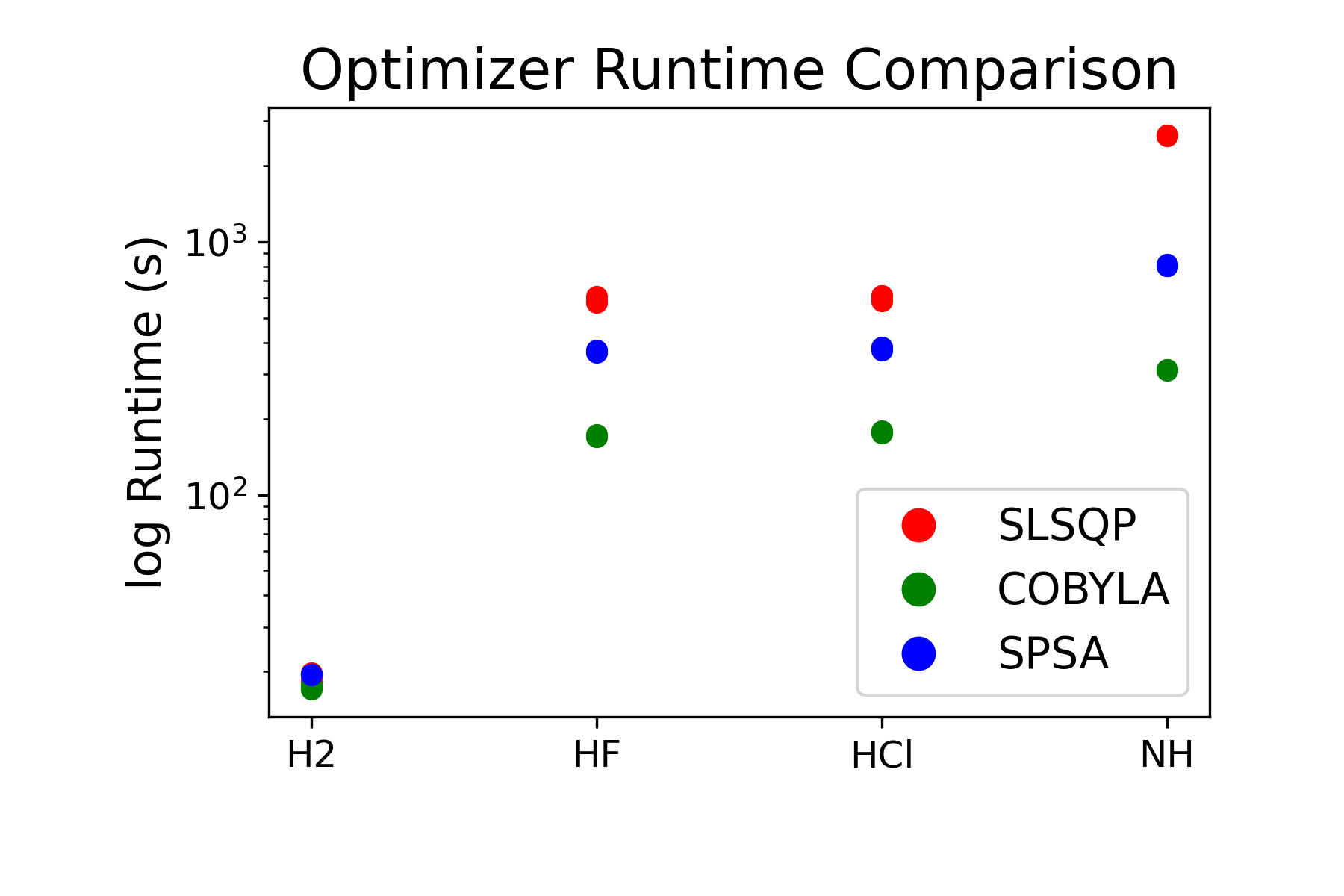}
            \caption{Runtime comparison in log scale for each optimizer-molecule combination.}
            \label{Runtime Comparison Plot}
        \end{center}
    \end{figure}

    These times are plotted in Figure \ref{Runtime Comparison Plot} for direct comparison across different optimizers and molecules.
    
    \pagebreak
    
    In every trial, COBYLA was the optimizer which completed the process the fastest. In all but two of the trials, SLSQP was the slowest optimizer and SPSA took second place. The two trials where this was different were trials $1$ and $3$ for H$_2$, where SLSQP was faster than SPSA by only $1.21$ and $1.24$ seconds respectively. Since the majority of the trials have the same speed rankings, we consider this to be generally true for simulations under these conditions.

\section{Discussion}

    The results of the accuracy and efficiency comparisons are not surprising when you consider how these optimizers work. In this chapter, we describe the inner workings of the optimizers and explain the best optimizer choices under different conditions.

\subsection{The Optimizers}

    SLSQP is an optimizer which can minimize a function of several variables containing any combination of bounds and constraints \cite{SLSQP}. This is ideal for problems where the objective function being optimized, as well as its constraints, are twice continuously differentiable. For values in the limit of infinity, large floating points values are used instead. SLSQP requires calculation of the objective function's second derivative, which takes a non-negligible amount of time. For this reason, the longer runtimes should be expected for the SLSQP optimizer.
    
    COBYLA is useful for constrained problems where the derivative of the objective function is unknown \cite{COBYLA}. The Qiskit textbook recommends this optimizer for noise-free objective functions when the minimum number of evaluations is desired. COBYLA only evaluates the objective function of the problem once per iteration, which makes it quite efficient. As a result, we see comparatively short runtimes for this optimizer.
    
    The SPSA optimizer estimates the gradient of the objective function after shifting the parameters in a random fashion. Qiskit recommends SPSA for noisy simulators or real hardware. Other works \cite{SPSA} agree that SPSA is ``most effective when the loss function measurements include added noise.'' On this non-noisy simulator, we see the effectiveness of the optimizer fade as evidenced by the variance between the energy curves produced by the code. SPSA evaluates the objective function twice, which explains why we see it perform slower than COBYLA in every trial. 

    Of course, these are not the only optimization methods that exist, but we wanted to ensure that the optimizers we examined were supported natively within Qiskit to avoid any compatibility issues. Some other optimizers, like the Rotosolve optimizer, have been shown to converge faster than some gradient-based optimizers. This optimizer ``does not allow for full potential for parallelization for VQE'' and requires three objective function samples, which is more than typical gradient descent methods \cite{Tilly_2022}. 

\subsection{Choosing an Optimizer}

    The best choice of optimizer ultimately comes down to what resources are available to the user and what they prioritize as most important. 
    
    For finding the ground state energies of simple molecules on a Qiskit simulator, the SLSQP optimizer determines the energy very well, obtaining results very close to an analytical classical eigensolver, and is easily the most accurate of the three tested. When time is in no short supply, SLSQP would be the best optimizer to choose.
    
    Applying the COBYLA optimizer to VQE to find the ground of state energies of molecules, this optimizer does well, but is not as accurate as SLSQP. This is likely due to the smaller number of function evaluations, causing a tradeoff between efficiency and accuracy. Still, it gave very good approximations for a majority of the atomic distances for each molecule and was much faster than SLSQP in several cases. When time is limited, COBYLA is the best choice of the three.
    
    For simulations on relatively low-end, non-noisy simulators, SPSA is not recommended. The results tend not to correlate well at all with the energies determined by classical means. Unfortunately, it is also not fast enough to make up for this loss of accuracy. When executing VQE on a noisy device, SPSA may have some gains over SLSQP and COBYLA, but we have yet to test this.
    
\section{Future Work}

    The timeframe allowed for this research has left some further work to be desired. Here, we suggest some ways to continue building on these findings.   
    
    The first suggestion would be to introduce more modifications to the code to improve its efficiency outside of the optimizer. Not every molecule worth examining will be small enough to allow for reasonably efficient computation on low-end hardware, and it is not feasible to wait many hours for a small number of steps to be calculated.
    
    Additionally, using improved hardware would also be a good way to continue this work. For these trials, the code was run on the University of Massachusetts Dartmouth Rapid Prototyping Server (RPS), which does not have particularly powerful specs available to a single end-user. Running this code on a more powerful device like a supercomputer could potentially allow for larger molecules to be simulated. Additionally, it would be expected that the runtimes would be reduced, allowing for adjustments and improvements to be made more rapidly.

    Finally, a deeper analysis of the experimental accuracy could be performed. In this work, we measured accuracy in terms of closeness between the VQE-calculated and classically-calculated energy curves. Comparing these results to the experimentally-determined energies of these molecules would give an improved description of the true accuracy of the process and its viability to outperform classical solvers. 









\bibliography{sources}{}
\end{document}